\def\PLUTO{{\sc pluto}}
\def\em{{\rm em}}
\def\rate{{\it \.M}}
\def\gsim{\;\lower4pt\hbox{${\buildrel\displaystyle >\over\sim}$}\,}

\newcommand\rs[1]{_\mathrm{#1}}

\newcommand{\referee}{ }

\documentclass{aa} 
\usepackage{natbib}
\usepackage{graphicx}
\usepackage{txfonts}
\begin{document}
   \title{X-ray emitting MHD accretion shocks in classical T Tauri stars}

   \subtitle{Case for moderate to high plasma-$\beta$ values}

   \author{S. Orlando\inst{1}
          \and
          G.G. Sacco\inst{2,}\inst{1}
          \and
          C. Argiroffi\inst{3,}\inst{1}
          \and
          F. Reale\inst{3,}\inst{1}
          \and
          G. Peres\inst{3,}\inst{1}
          \and
          A. Maggio\inst{1}
          }

   \institute{INAF-Osservatorio Astronomico di Palermo,
           Piazza del Parlamento, 1 , 90134, Palermo, Italy
           \email{orlando@astropa.inaf.it}
         \and
           Chester F. Carlson Center for Imaging Science, Rochester
           Inst. of Technology, 54 Lomb Memorial Dr., Rochester,
           NY, 14623, USA
         \and
           DSFA-Universit\`a degli Studi di Palermo, Piazza del
           Parlamento, 1, 90134, Palermo, Italy
 }

   \date{}

 
  \abstract
   {Plasma accreting onto classical T Tauri stars (CTTS) is believed to
    impact the stellar surface at free-fall velocities, generating
    a shock. Current time-dependent models describing accretion shocks
    in CTTSs are one-dimensional, assuming that the plasma moves and
    transport energy only along magnetic field lines ($\beta \ll
    1$).}
   {We investigate the stability and dynamics of accretion shocks in
    CTTSs, considering the case of $\beta \gsim 1$ in the post-shock
    region. In these cases the 1D approximation is not valid and a
    multi-dimensional MHD approach is necessary.}
   {We model an accretion stream propagating through the atmosphere of
    a CTTS and impacting onto its chromosphere, by performing 2D
    axisymmetric MHD simulations. The model takes into account the
    stellar magnetic field, the gravity, the radiative cooling, and the
    thermal conduction (including the effects of heat flux saturation).}
   {The dynamics and stability of the accretion shock strongly depends
    on the plasma $\beta$. In the case of shocks with $\beta > 10$,
    violent outflows of shock-heated material (and possibly MHD waves)
    are generated at the base of the accretion column and strongly perturb
    the surrounding stellar atmosphere and the accretion column itself
    (modifying, therefore, the dynamics of the shock). In shocks with
    $\beta \approx 1$, the post-shock region is efficiently confined
    by the magnetic field. The shock oscillations induced by cooling
    instability are strongly influenced by $\beta$: for $\beta > 10$,
    the oscillations may be rapidly dumped by the magnetic field,
    approaching a quasi-stationary state, or may be chaotic with no
    obvious periodicity due to perturbation of the stream induced by the
    post-shock plasma itself; for $\beta\approx 1$ the oscillations are
    quasi-periodic, although their amplitude is smaller and the frequency
    higher than those predicted by 1D models.}
   {}

   \keywords{accretion, accretion disks --
             Instabilities --
             Magnetohydrodynamics (MHD) --
             Shock waves --
             stars: pre-main sequence --
             X-rays: stars}

   \titlerunning{X-ray emitting MHD accretion shocks in CTTSs}

   \maketitle
%

\section{Introduction}

In the last few years, high resolution ($R=600$) X-ray observations of
\referee{several} young stars accreting material from their circumstellar
disk (TW~Hya, BP~Tau, V4046~Sgr, MP~Mus and RU~Lupi) suggested the
presence of X-ray emission at temperature $T=2-5$ MK from plasma
denser than $n_{\rm H} = 10^{11}$ cm$^{-3}$ (\citealt{Kastner2002ApJ,
Schmitt2005A&A, Gunther2006A&A, Argiroffi2007A&A, Robrade2007A&A,
2009A&A...507..939A}). The emitting plasma is too dense to be enclosed
inside coronal loop structures (\citealt{2004ApJ...617..508T}). Several
authors suggested, therefore, that this soft X-ray emission component
could be produced by the material accreting onto the star surface,
flowing along the magnetic field lines of the nearly dipolar stellar
magnetosphere, and heated to temperatures of few MK by a shock at the
base of the accretion column \citep{Calvet1998ApJ, Lamzin1998ARep}.

The idea that an accreting flow impacting onto a stellar surface
leads to a shocked slab of material heated at millions degrees is not
new. For instance, in the context of degenerate stars, several authors
have studied the dynamics and energetic of accretion shocks and the
effects of radiation on the formation, structure, and stability of the
shocks \citep{Langer1981ApJ, 1982ApJ...258..289L, Chevalier1982ApJ,
1985ApJ...296..128I, 1985ApJ...299L..87C}. \cite{1998A&A...340..593H}
have also investigated the role of the magnetic field in confining the
post-shock accreting material and in determining the evolution of the
accretion shock. These studies have shown that the role of thermal and
dynamical instabilities is critical to our ability to model radiative
shock waves.

The first attempt to interpret the evidence of soft X-ray emission from
dense plasma in Classical T Tauri stars (CTTSs) in terms of accretion
shocks has quantitatively demonstrated that some non coronal features
of the X-ray observations of TW~Hya \citep{Gunther2007A&A} and MP~Mus
\citep{Argiroffi2007A&A} could be explained through a simplified
steady-flow shock model. However, steady models are known to be not a
good approximation for radiative shocks, since they neglect the
important effects of local thermal instabilities as well as global shock
oscillations induced by radiative cooling \citep{1993ApJS...88..253S,
1996ApJS..102..161D, 1998ApJ...494..336S, 2003ApJ...591..238S,
2003ApJS..147..187S, 2005ApJ...626..373M}.

In fact, the first time-dependent 1D models of radiative accretion shocks
in CTTSs have shown quasi-periodic oscillations of the shock position
induced by cooling \citep{Koldoba2008MNRAS, 2008A&A...491L..17S}. In
particular, \cite{2008A&A...491L..17S} have developed a 1D hydrodynamic
model including, among other important physical effects, a well tested
detailed description of the stellar chromosphere and investigated the
role of the chromosphere in determining the position and the thickness
of the shocked region. \referee{For hydrodynamical simulations} based on the
parameters of MP~Mus, \cite{2008A&A...491L..17S} synthesized the high
resolution X-ray spectrum, as it would be observed with the Reflection
Grating Spectrometers (RGS) on board the XMM-Newton satellite, and found
an excellent agreement with the observations.

Up to date time-dependent models of accretion shocks in CTTSs have been
one dimensional, assuming that the plasma moves and transports energy only
along magnetic field lines. This hypothesis is justified if the plasma
$\beta \ll 1$ (where $\beta =$ gas pressure / magnetic pressure) in the
shock-heated material. The photospheric magnetic field magnitude of CTTSs
is believed to be around 1 kG (e.g. \citealt{1999ApJ...510L..41J}). Such
a strong stellar field is enough to confine efficiently accretion shocks
with particle density below $10^{13}$ cm$^{-3}$ and temperature around
5 MK ($\beta < 0.3$). However, recent polarimetric measurements indicate
that, in some cases, the photospheric magnetic field strength could be
less than 200 G \citep{2004Ap&SS.292..619V} and the plasma $\beta$ in
the slab may be around 1 or even larger. In these cases, the magnetic
field configuration in the post-shock region may change, influencing the
physical structure of the material emitting in X-rays and the stability
of the accretion shock.

The low-$\beta$ approximation was challenged by recent findings of
\cite{2009ApJ...703.1224D}. In case of $\beta \ll 1$, the radiative
shock instability is expected to lead to detectable periodic modulation
of the X-ray emission from the shock-heated plasma, if the density and
velocity of the accretion stream do not change over the time interval
considered (in agreement with predictions of 1D models). However, the
analysis of soft X-ray observations of the CTTS TW~Hya (whose emission
is believed to arise predominantly from accretion shocks) produced no
evident periodic variations (\citealt{2009ApJ...703.1224D}). These authors
concluded, therefore, that 1D models may be too simple to describe the
multi-dimensional structure of the shock and that the magnetic field may
play an important role through the generation and damping of MHD waves.

In this paper, we investigate the stability and dynamics of accretion
shocks in cases for which the low-$\beta$ approximation cannot be
applied. We analyze the role of the stellar magnetic field in the
dynamics and confinement of the slab of shock-heated material. To this
end, we model an accretion stream propagating through the magnetized
atmosphere of a CTTS and impacting onto its chromosphere, using 2D
axisymmetric MHD simulations and, therefore, an explicit description of
the ambient magnetic field. We investigate cases of $\beta \gsim 1$ in
the post-shock region for which the deviations from 1D models are expected
to be the largest. In Sect. \ref{sec2} we describe the MHD model, and
the numerical setup; in Sect. \ref{sec3} we describe the results;
in Sect. \ref{sec4} we discuss the implications of our results
and, finally, in Sect. \ref{sec5}, we draw our conclusions.

\section{MHD modeling}
\label{sec2}

The model describes an accretion stream impacting onto the surface of
a CTTS. We assume that the accretion occurs along magnetic field lines
linking the circumstellar disk to the star and consider only the portion
of the stellar atmosphere close to the star.

\begin{figure}[!t]
  \centering
  \includegraphics[width=8.5cm]{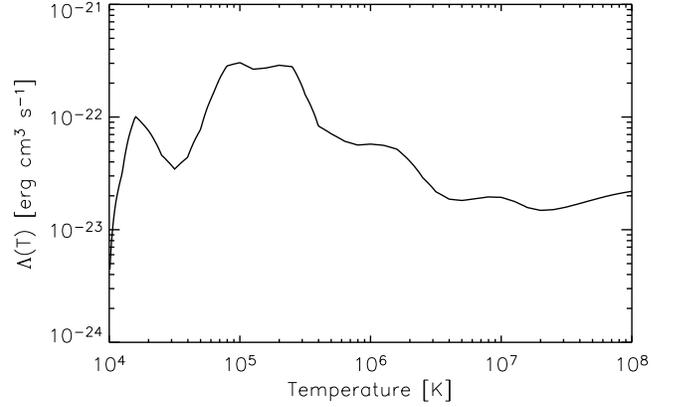}
  \caption{Radiative losses for an optically thin plasma from the APED
           V1.3 atomic line database \citep{Smith2001ApJ}, assuming
           the metal abundances 0.5 the solar values
           (\citealt{Anders1989GeCoA}).}
  \label{rad_loss}
\end{figure}

The fluid is assumed to be fully ionized with a ratio of specific heats
$\gamma = 5/3$. The model takes into account the stellar magnetic field,
the gravity, the radiative cooling, and the thermal conduction (including
the effects of heat flux saturation). \referee{Since the magnetic Reynolds
number $\gg 1$ considering the typical velocity ($10^7$ cm s$^{-1}$) and
length scale ($10^9$ cm) of the system, the flow is treated as an ideal
MHD plasma.} The impact of the accretion stream is modeled by solving
numerically the time-dependent MHD equations (written in non-dimensional
conservative form):

\begin{equation}
\frac{\partial \rho}{\partial t} + \nabla \cdot (\rho \vec{u}) = 0~,
\end{equation}

\begin{equation}
\frac{\partial \rho \vec{u}}{\partial t} + \nabla \cdot (\rho
\vec{u}\vec{u}-\vec{B}\vec{B}) + \nabla P_* = \rho \vec{g}~,
\end{equation}

\begin{eqnarray}
\lefteqn{\frac{\partial \rho E}{\partial t} +\nabla\cdot [\vec{u}(\rho
E+P_*) -\vec{B}(\vec{u}\cdot \vec{B})] =} \nonumber \\
 & \displaystyle ~~~~~~~~~~~~~~~~~~~~~~~~~ \rho \vec{u}\cdot
\vec{g} -\nabla\cdot \vec{F}_{\rm c} -n_{\rm e} n_{\rm H}
\Lambda(T)~,
\end{eqnarray}

\begin{equation}
\frac{\partial \vec{B}}{\partial t} +\nabla
\cdot(\vec{u}\vec{B}-\vec{B}\vec{u}) = 0~,
\end{equation}

\noindent
where

\[
P_* = P + \frac{B^2}{2}~,~~~~~~~~~~~~~
E = \epsilon +\frac{1}{2} u^2+\frac{1}{2}\frac{B^2}{\rho}~,
\]

\noindent
are the total pressure, and the total gas energy (internal energy,
$\epsilon$, kinetic energy, and magnetic energy) respectively, $t$
is the time, $\rho = \mu m_H n_{\rm H}$ is the mass density, $\mu =
1.28$ is the mean atomic mass (assuming metal abundances 0.5 the solar
values; \citealt{Anders1989GeCoA}), $m_H$ is the mass of the hydrogen
atom, $n_{\rm H}$ is the hydrogen number density, $\vec{u}$ is the
gas velocity, $g$ is the gravity, $T$ is the temperature, $\vec{B}$
is the magnetic field, $\vec{F}_{\rm c}$ is the conductive flux, and
$\Lambda(T)$ represents the \referee{optically thin} radiative losses per
unit emission measure\footnote{\referee{Note that the radiative losses are
dominated by emission lines in the temperature regime common to CTTSs and
are set equal to zero for $T< 10^4$ K.}} \referee{(see Fig.~\ref{rad_loss})}
derived with the PINTofALE spectral code \citep{Kashyap2000BASI} with the
APED V1.3 atomic line database \citep{Smith2001ApJ}, assuming the same
metal abundances as before (as deduced from X-ray observations of CTTSs;
\citealt{Telleschi2007A&Ab}). We use the ideal gas law, $P=(\gamma-1)
\rho \epsilon$.

The thermal conductivity in an organized magnetic field is known to
be highly anisotropic and it can be highly reduced in the direction
transverse to the field. The thermal flux, therefore, is locally
split into two components, along and across the magnetic field lines,
$\vec{F}_{\rm c} = F_{\parallel}~\vec{i}+F_{\perp}~\vec{j}$, where (see,
for instance, \citealt{2008ApJ...678..274O})

\begin{equation}
\begin{array}{l}\displaystyle
F_{\parallel} = \left(\frac{1}{[q_{\rm spi}]_{\parallel}}+
                \frac{1}{[q_{\rm sat}]_{\parallel}}\right)^{-1}~,
\\ \\ \displaystyle
F_{\perp} = \left(\frac{1}{[q_{\rm spi}]_{\perp}}+
               \frac{1}{[q_{\rm sat}]_{\perp}}\right)^{-1}~,
\end{array}
\label{cond}
\end{equation}

\noindent
to allow for a smooth transition between the classical and saturated
conduction regime. In Eqs. \ref{cond}, $[q_{\rm spi}]_{\parallel}$ and
$[q_{\rm spi}]_{\perp}$ represent the classical conductive flux along
and across the magnetic field lines \citep{spi62}

\begin{equation}
\begin{array}{l}\displaystyle
[q_{\rm spi}]_{\parallel} = -\kappa_{\parallel} [\nabla T]_{\parallel}
\approx - 9.2\times 10^{-7} T^{5/2}~ [\nabla T]_{\parallel}
\\ \\ \displaystyle
[q_{\rm spi}]_{\perp} = -\kappa_{\perp} [\nabla T]_{\perp}
\approx - 3.3\times 10^{-16} \frac{n^2_{\rm H}}{T^{1/2}B^2}~ [\nabla
T]_{\perp}
\end{array}
\label{spit_eq}
\end{equation}

\noindent
where $[\nabla T]_{\parallel}$ and $[\nabla T]_{\perp}$ are the thermal
gradients along and across the magnetic field, and $\kappa_{\parallel}$
and $\kappa_{\perp}$ (in units of erg s$^{-1}$ K$^{-1}$ cm$^{-1}$) are
the thermal conduction coefficients along and across the magnetic field
lines, respectively. The saturated flux along and across the magnetic
field lines, $[q_{\rm sat}]_{\parallel}$ and $[q_{\rm sat}]_{\perp}$,
are \citep{cm77}

\begin{equation}
\begin{array}{l}\displaystyle
[q_{\rm sat}]_{\parallel} = -\mbox{sign}\left([\nabla
T]_{\parallel}\right)~
                5\phi \rho c_{\rm s}^3,
\\ \\ \displaystyle
[q_{\rm sat}]_{\perp} = -\mbox{sign}\left([\nabla T]_{\perp}\right)~
                5\phi \rho c_{\rm s}^3,
\end{array}
\label{therm}
\end{equation}

\noindent
where $c_{\rm s}$ is the isothermal sound speed, and $\phi$ is a number
of the order of unity; we set $\phi = 1$ according to the values suggested
for stellar coronae \citep[][ and references therein]{1984ApJ...277..605G,
1989ApJ...336..979B, 2002A&A...392..735F}.

\begin{figure}[!t]
  \centering
  \includegraphics[width=8.cm]{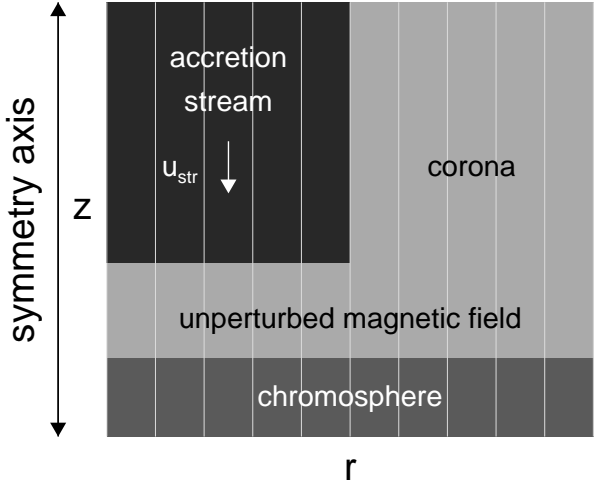}
  \caption{Initial geometry of the system in cylindrical coordinates. The
           stellar surface lies on the $r$ axis and the unperturbed
           stellar magnetic field is uniform and oriented along the $z$
           axis (vertical lines). The accretion stream propagates
           downwards through the stellar corona with velocity
           $u\rs{str}$. The $z$-axis is the symmetry axis of the problem.}
  \label{fig1}
\end{figure}

\begin{figure}[!t]
  \centering
  \includegraphics[width=8.5cm]{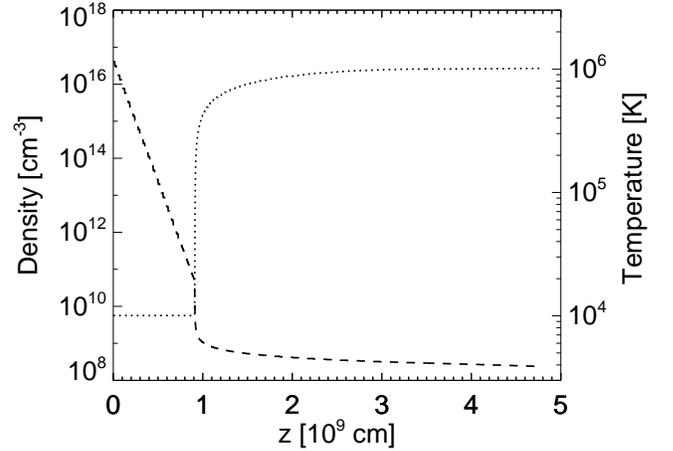}
  \caption{Initial hydrogen number density (dashed line) and
      temperature (dotted line) as a function of height above the stellar
      surface, $z$, for the unperturbed stellar atmosphere.}
  \label{fig2}
\end{figure}

We solve the MHD equations using cylindrical coordinates in the plane
$(r,z)$, assuming axisymmetry and the stellar surface lying on the $r$
axis (see Fig.~\ref{fig1}). The initial unperturbed stellar atmosphere
is assumed \referee{to be magneto-static} and to consist of a hot (maximum
temperature $\approx 10^6$ K) and tenuous ($n_{\rm H} \approx 2\times
10^8$ cm$^{-3}$) corona linked through a steep transition region to
an isothermal chromosphere\footnote{Note that the radiative losses
are set to zero in the chromosphere to keep it in equilibrium.} that
is uniformly at temperature $10^4$ K and is $8.5\times 10^8$ cm thick
(see Fig.~\ref{fig2}). The choice of an isothermal chromosphere is
adopted for ease of modeling, and different choices of more detailed
chromospheric models have been shown not to lead to significantly
different results (Sacco et al. 2009, in preparation). The unperturbed
stellar magnetic field is uniform, oriented along the $z$ \referee{axis
and perpendicular} to the stellar surface. The gravity is calculated
assuming the star mass $M=1.2 M_{\sun}$ and the star radius $R=1.3
R_{\sun}$ \referee{which is} appropriate for the CTTS MP~Mus (see
\citealt{Argiroffi2007A&A}). Different choices of stellar parameters
should not lead to different results.

\referee{The accretion stream is assumed to be constant and propagates along
the $z$ axis. Initially the stream is in pressure equilibrium with the
stellar corona and has a circular cross-section} with radius $r\rs{str}
= 5\times 10^{9}$ cm; its radial density distribution is given by

\begin{equation}
n\rs{str}(r) = n_{\rm cor}+\frac{n_{\rm str0}-n_{\rm cor}}
{\cosh\left[\sigma\left(r/r_{\rm str}\right)^{\sigma}\right]}~~~,
\end{equation}

\noindent
where $n\rs{str0}$ and $n_{\rm cor}$ are the hydrogen number density
at the stream center and in the corona, and $\sigma=20$ is a steepness
parameter. The above distribution describes a thin shear layer ($\sim
0.1~r_{\rm str}$) around the stream that smooths the stream density to
the value of the surrounding medium and maintains numerical stability,
allowing for a smooth transition of the Alfv\'en speed at the interface
between the stream and the stellar atmosphere. The stream temperature
is determined by the pressure balance across the stream boundary.

The velocity of the stream along the $z$ axis also has a radial
profile:

\begin{equation}
u\rs{str}(r) = \frac{u_{\rm str0}}
{\cosh\left[\sigma\left(r/r_{\rm str}\right)^{\sigma}\right]}~~~,
\end{equation}

\noindent
where $u\rs{str0}$ is the free fall velocity at the center of the
stream. The smooth shear layer of velocity avoids the growth of
random perturbations with wavelengths of the order of the grid size
(e.g. \citealt{1994A&A...283..655B}).

The simulations presented here describe the evolution of the system
for a time interval of about 3000 s. We used the accretion parameters
(velocity and density) that match the soft X-ray emission of MP~Mus
(\citealt{Argiroffi2007A&A}), namely $n_{\rm str0} = 10^{11}$ cm$^{-3}$
and $u_{\rm str0} = -500$ km s$^{-1}$ at a height $z =2.4\times
10^{10}$~cm above the stellar surface\footnote{Note that the stream
velocity is $u_{\rm str0} = -580$ km s$^{-1}$ when the stream impacts
onto the chromosphere, due to gravity.}. The initial field strengths of
$|\vec{B}\rs{0}| = 1,\,10,\,50$~G in the unperturbed stellar atmosphere
(see Table \ref{tab1}) maintain $\beta \gsim 1$ in the post-shock region,
where $\beta=P/(B^2/8\pi)$ is the ratio of thermal to magnetic
pressure. For instance, $|\vec{B}\rs{0}| = 10$~G corresponds to $\beta$
ranging between $\approx 10^4$, at the base of the chromosphere, and
$\approx 10^{-2}$ up in corona. We also performed a 1D hydrodynamic
simulation in order to provide a baseline for the 2D calculations. This
simulation corresponds to the strong magnetic field limit ($\beta \ll
1$), in which the plasma moves and transports energy exclusively along
the magnetic field lines (see \citealt{2008A&A...491L..17S}).

\begin{table}
\centering
\caption{Relevant parameters of the simulations}
\label{tab1}
\begin{tabular}{lcccc}
\hline
\hline
Name of    &  $|\vec{B}\rs{0}|$ [G] & Geometry & Mesh & time \\
simulation &                        &          &      & covered [s] \\
\hline 
HD-1D    & -            & Cartes. 1D & 1024 & 3000 \\
By-01    & 1            & cylind. 2D & $512\times 1024$  & 3000 \\
By-10    & 10           & cylind. 2D & $512\times 1024$  & 4000 \\
By-50    & 50           & cylind. 2D & $512\times 1024$  & 3000 \\
By-01-HR & 1            & cylind. 2D & $1024\times 2048$ & 1000 \\
By-10-HR & 10           & cylind. 2D & $1024\times 2048$ & 1000 \\
By-50-HR & 50           & cylind. 2D & $1024\times 2048$ & 2000 \\
\hline
\hline
\end{tabular}
\end{table}

The calculations described in this paper were performed using
\PLUTO\ \citep{2007ApJS..170..228M}, a modular, Godunov-type code for
astrophysical plasmas. The code provides a multiphysics, multialgorithm
modular environment particularly oriented toward the treatment of
astrophysical flows in the presence of discontinuities such as in the
case treated here. \referee{The code was designed to make efficient
use of massively parallel computers using the message-passing interface
(MPI) library for interprocessor communications. The MHD equations are solved
using the MHD module available in \PLUTO\ which is based on the
Harten-Lax-van Leer Discontinuities (HLLD) approximate Riemann solver
(\citealt{2005JCoPh.208..315M}). \cite{2005JCoPh.208..315M} have shown
that the HLLD algorithm can exactly solve isolated discontinuities
formed in the MHD system and, therefore, the adopted scheme is
particularly appropriate to describe accretion shocks. The evolution
of the magnetic field is carried out using the constrained transport
method of \cite{1999JCoPh.149..270B} that maintains the solenoidal
condition at machine accuracy. \PLUTO\ includes optically thin radiative
losses in a fractional step formalism \citep{2007ApJS..170..228M}, which
preserve $2^{nd}$ time accuracy, being the advection and source steps at
least $2^{nd}$ order accurate; the radiative losses $\Lambda$ values are
computed at the temperature of interest using a table lookup/interpolation
method. The thermal conduction is solved with an explicit scheme that
adds the parabolic contributions to the upwind fluxes computed at cell
interfaces \citep{2007ApJS..170..228M}. Such a scheme is subject to
a rather restrictive stability condition (i.e. $\Delta t < (\Delta
x)^2/(2\eta)$, where $\eta$ is the maximum diffusion coefficient),
being the thermal conduction timescale generally shorter than the
dynamical one (e.g. \citealt{2000A&A...362L..41H, 2005CoPhC.168....1H,
2005A&A...444..505O, 2008ApJ...678..274O}).}

The symmetry of the problem allows us to solve the MHD equations in
half of the spatial domain with the stream axis coincident with the $z$
axis. The 2D cylindrical $(r,z)$ mesh extends between 0 and $1.2 \times
10^{10}$ cm in the $r$ direction and between 0 and $2.4 \times 10^{10}$
cm in the $z$ direction and consists of a uniform grid with $512\times
1024$ grid points. Additional runs were done with setups identical to
those used for runs By-01, By-10, and By-50 but with higher resolution
($1024\times 2048$ grid points; see Tab. \ref{tab1}) to evaluate the
effect of spatial resolution (see Sect. \ref{resol}); the simulations
with higher resolution cover a time interval of about 1000 s.

We use fixed boundary conditions at the lower ($z = 0$) boundary,
\referee{imposing zero material and heat flux across the boundary. With
this condition, matter may progressively accumulate at the base of the
chromosphere; we have estimated that this effect may become significant
on timescales a factor 100 longer than than those explored by our
simulations.} Axisymmetric boundary conditions\footnote{\referee{Variables
are symmetrized across the boundary and normal components and angular
$\phi$ components of vector fields ($\vec{u}, \vec{B}$) flip signs.}}
are imposed at $r = 0$ (i.e. along the symmetry axis of the problem),
a constant inflow at the upper boundary ($z = 2.4 \times 10^{10}$ cm)
for $r \leq r_{\rm str}$, and free outflow\footnote{\referee{Set zero
gradients across the boundary.}} elsewhere.

\section{Results}
\label{sec3}

\subsection{One-dimensional reference model}

The results of our 1D reference hydrodynamic simulation, HD-1D, are
analogous to those discussed by \cite{2008A&A...491L..17S}: the impact of
the accretion stream onto the stellar chromosphere generates a reverse
shock which propagates through the accretion column, producing a hot
slab. According to Sacco et al., the expected temperature, $T\rs{slab}$,
and maximum thickness, $D\rs{slab}$, of the slab, in the strong shock
limit (\citealt{ZelDovich1967book}) are:

\begin{equation}
T\rs{slab} \approx \frac{3}{32}\frac{\mu
m\rs{H}}{k\rs{B}}u\rs{str0}^2\approx 5~\mbox{MK}~,
\label{temperat}
\end{equation}

\begin{equation}
D\rs{slab} \approx 4.2\times 10^2
\frac{u\rs{str0}}{n\rs{str0}} T\rs{slab}^{3/2}\approx 2.5\times
10^9~\mbox{cm}~,
\label{thickness}
\end{equation}

\noindent
where $k\rs{B}$ is the Boltzmann constant.

\begin{figure}[!t]
  \centering
  \includegraphics[width=8.5cm]{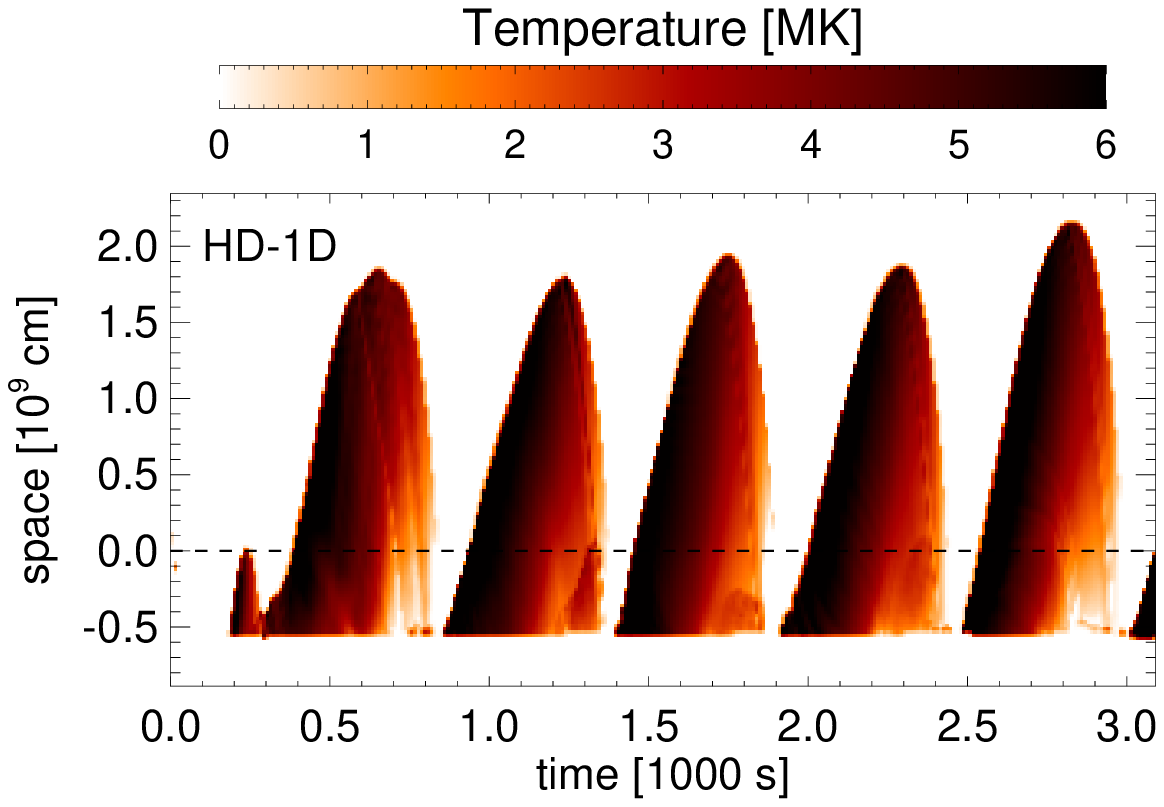}
  \caption{Time-space plot of the temperature evolution for run HD-1D. The
           spatial extent of the shock lies in the vertical direction
           at any instant in time. The dashed line marks the initial
           position of the transition region between the chromosphere
           and the corona.}
  \label{fig3}
\end{figure}

Fig. \ref{fig3} shows the time-space plot of the temperature evolution
for run HD-1D. The base of the hot slab penetrates the chromosphere
(the dashed line in Fig.~\ref{fig3} marks the initial position of the
transition region between the chromosphere and the corona) down to
the position at which the ram pressure, $P\rs{ram} = \rho\rs{str0}
u\rs{str0}^2$, of the post-shock plasma equals the thermal pressure
of the chromosphere (\citealt{2008A&A...491L..17S}). As evident from
the figure, the shock front is not steady and the amplitude and period of
the pulses rapidly reach stationary values as the initial transient
disappears. The shock position oscillates with a period of $\approx
600$~s due to strong radiative cooling at the base of the slab, which
robs the post-shock plasma of pressure support, causing the material
above the cooled layer to collapse back before the slab expands
again (see \citealt{2008A&A...491L..17S} for a detailed description
of the system evolution). The maximum thickness of the slab is
$D\rs{slab}\approx 2.5\times 10^9$~cm, in agreement with the prediction
(see Eq.~\ref{thickness}). The post-shock plasma reaches the temperature
$T\rs{slab} \approx 5-7$ MK during the expansion phase, and $T\rs{slab}
\approx 2-4$ MK during the cooling phase.

As already mentioned, the evolution of the accretion shock described
by the 1D reference model is appropriate if the magnetic field lines
can be considered rigid to any force exerted by the accreting plasma
flowing along the lines ($\beta \ll 1$). In case of plasma $\beta$ values
around 1 or even higher, the hot slab is expected to be only partially
confined by the magnetic field and flow can occur sideways because of
the strong pressure of the post-shock plasma and may ultimately perturb
the dynamics of the shock itself. Moreover, 2D or 3D structures leading
to strongly cooling zones are expected to develop in the post-shock
plasma if the latter is characterized by $\beta \gsim 1$. In this case,
\cite{2003ApJ...591..238S} showed that the evolution of 1D and 2D
radiative shocks may be significantly different.

\subsection{Two-dimensional radiative MHD shock model}

The cases of $\beta \gsim 1$ are described by our 2D model. The main
differences on the evolution from the 1D model are expected to occur
at the border of the stream where the shock-heated plasma might be not
confined efficiently by the magnetic field; so we focus our attention
there.

\begin{figure*}[!t]
  \centering
  \includegraphics[width=16.5cm]{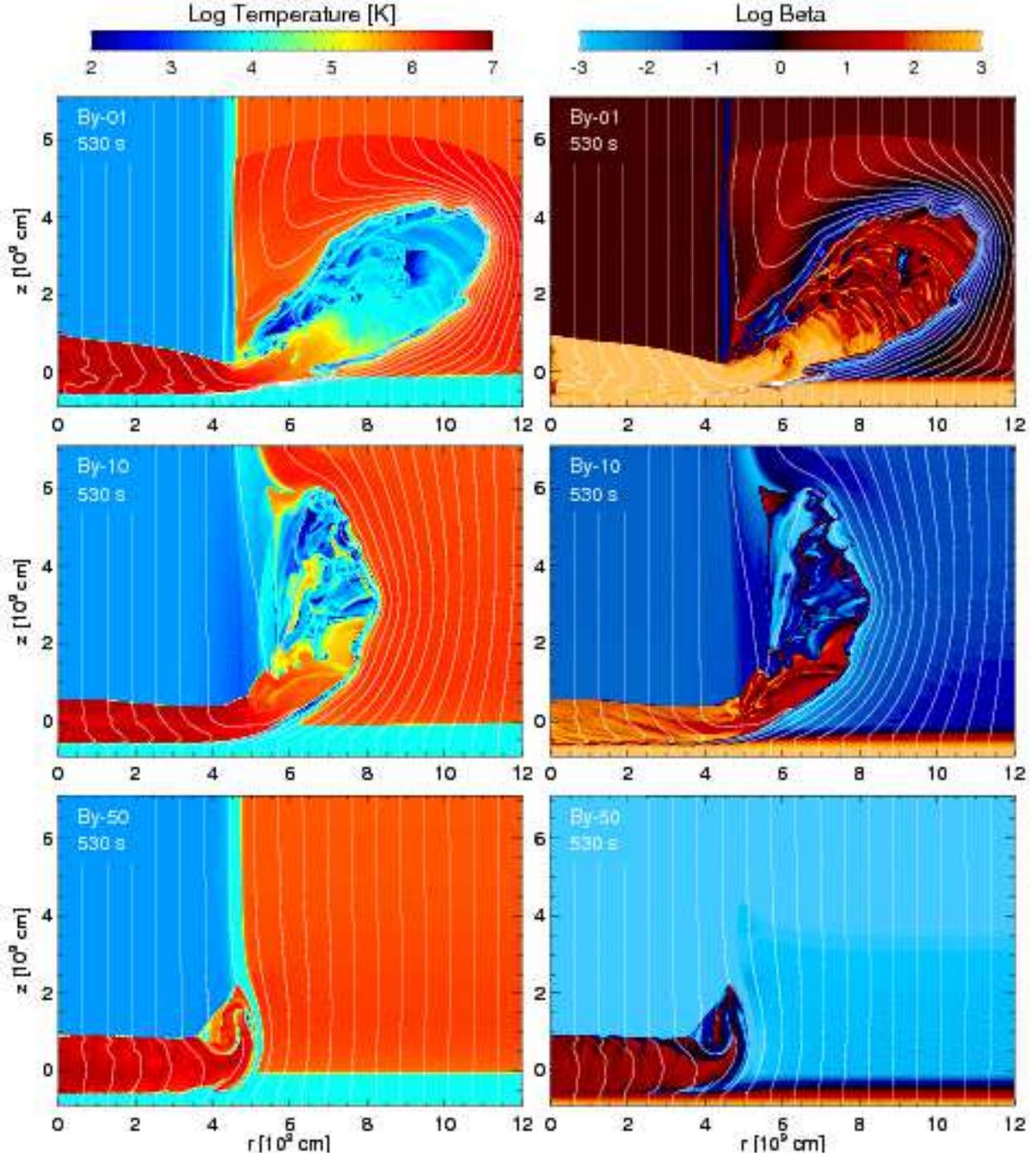}
  \caption{Temperature (left panels) and plasma $\beta$ (right panels)
   distributions in the $(r,z)$ plane, in log scale, in the simulations
   By-01, By-10, and By-50 at time $t=530$~s. The initial position of
   the transition
   region between the chromosphere and the corona is at $z=0$.
   The magnetic field is initially oriented along the $z$ axis; the
   white lines mark magnetic field lines.}
  \label{fig4}
\end{figure*}

Fig.~\ref{fig4} shows the spatial distribution of temperature (on the
left) and plasma $\beta$ (on the right) for runs By-01, By-10, and By-50
at time $t=530$~s (at early stage of evolution). Movies showing the
complete evolution of 2D spatial distributions of mass density (on the
left) and temperature (on the right), in log scale, for runs By-01, By-10,
and By-50 are provided as on-line material. The accretion flow follows
the magnetic field lines and impacts onto the chromosphere, forming a hot
slab at the base of the stream with temperatures $\approx 5$ MK and $\beta
\gsim 1$. In all the cases, the slab is rooted down in the chromosphere,
where the thermal pressure equals the ram pressure, and part ($\approx
1/3$) of the shock column is buried under a column of optically thick
material and may suffer significant absorption. In runs By-01 and By-10,
the dense hot plasma behind the shock front causes a pressure driven flow
parallel to the stellar surface, expelling accreted material sideways
(see upper and middle panels in Fig.~\ref{fig4}; see also on-line movies
for runs By-01 and By-10). This outflow strongly perturbs the shock
dynamics and is absent in the 1D shock model. As expected, this feature
determines the main differences between these 2D and the 1D simulations.

In run By-01, the magnetic field is too weak to confine the post-shock
plasma ($\beta \approx 10^4$ at the border of the slab) and a
conspicuous amount of material continuously escapes from the border of
the accretion column at its base where the flow impacts onto the stellar
surface. The maximum escape velocity is comparable to the free-fall
velocity, $u\rs{str0}$, and this outflow acts as an additional cooling
mechanism. The resulting outflow advects and stretches the magnetic field
lines (see upper panels in Fig.~\ref{fig4}), taking the material away
from the accretion column and strongly perturbing the stellar atmosphere
even at several stream radii. As a result of the outflow, a strong
component of $\vec{B}$ perpendicular to the stream velocity ($B\rs{r}
\approx 0.3 |\vec{B}\rs{0}|$) appears in the post-shock region (see
upper panels in Fig.~\ref{fig4}). As discussed by
\cite{1993ApJ...413..176T}, the presence of even a small magnetic field
perpendicular to the flow can stabilize the overstable oscillations. In
fact, at variance with the force due to gas pressure, the Lorentz force is
not affected by cooling processes (see also \citealt{1998A&A...340..593H})
and this mechanism may contribute to stabilize the shock oscillations
(see the on-line movie).

\begin{figure*}[!th]
  \centering
  \includegraphics[width=16.5cm]{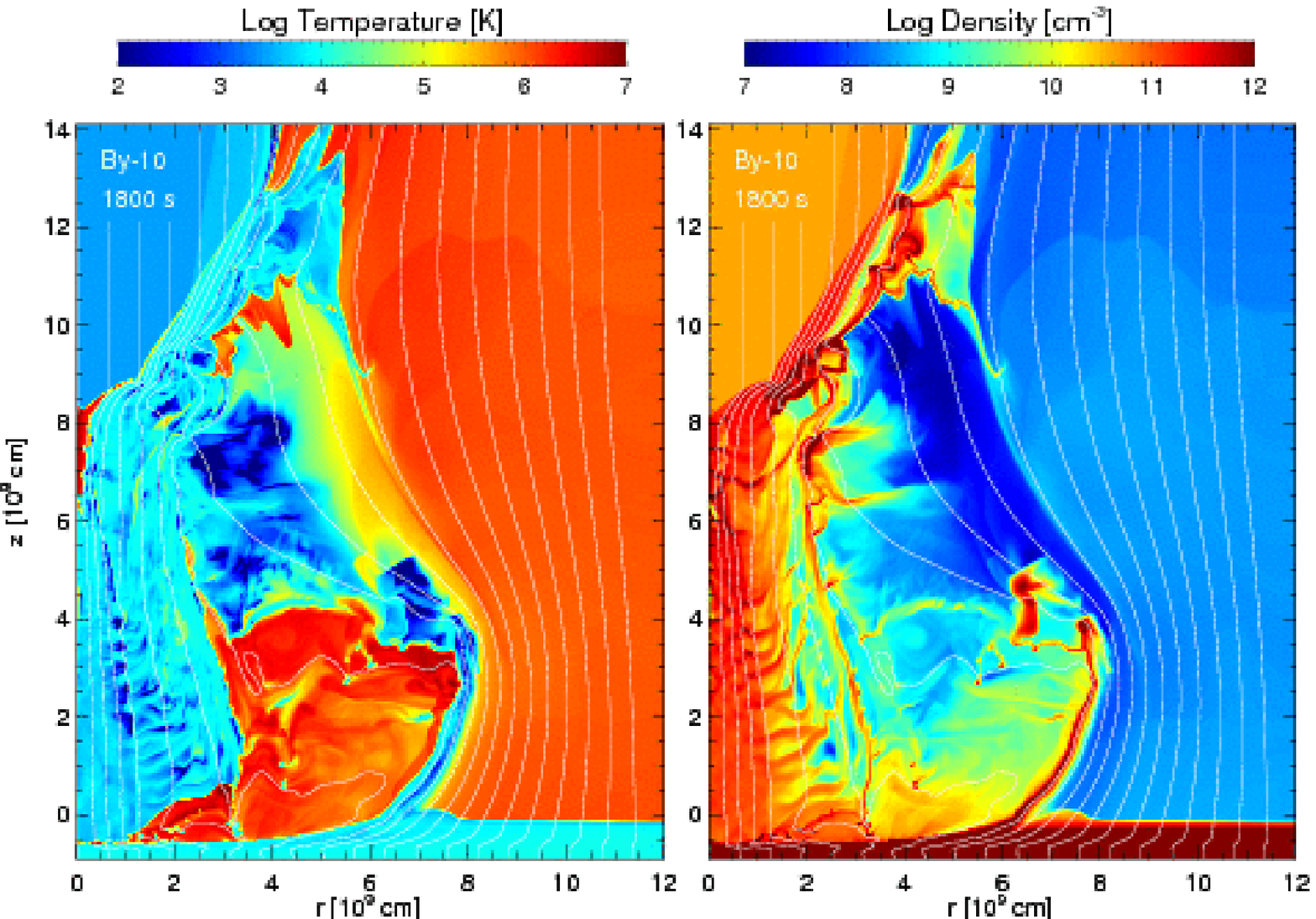}
  \caption{Temperature (left panel) and mass density (right panel)
   distributions in the $(r,z)$ plane, in log scale, in the simulation
   By-10, at time $t=1800$~s. The white lines mark magnetic
   field lines.}
  \label{fig5}
\end{figure*}

In the intermediate run By-10, the magnetic field is trapped at the head
of the escaped material, leading to a continuous increase of the magnetic
pressure and field tension there. As a result, the escaped material is
kept close to the accretion column by the magnetic field (at variance
with the By-01 case) and, eventually, may perturb the stream itself. In
fact, the escaped material accumulates around the accretion column,
forming a growing sheath of turbulent material gradually enveloping
the stream. The magnetic pressure and field tension increase at the
interface between the ejected material and the surrounding medium,
pushing on the material and forcing it to plunge into the stream after
$\approx 1.3$~ks. Fig.~\ref{fig5} shows the temperature and mass density
distributions at time $t=1.8$~ks, when the expelled material has already
entered into the stream and has strongly perturbed the accretion column
(see also the on-line movie to follow the complete evolution). As
a result of the stream perturbation, the hot slab may temporarily
disappear altogether as shown in Fig.~\ref{fig5}. In this phase, the
region of impact of the stream onto the chromosphere is characterized
by a rather complex structure with knots and filaments of material and
may involve possible mixing of plasma of the surrounding corona with
accretion material.

In the By-50 case, the post-shock plasma is confined efficiently by the
magnetic field and no outflow of accreted material forms (see lower
panels in Fig.~\ref{fig4}). In particular, $\beta \approx 10^{-4}$
and the magnetic pressure $P\rs{B} \approx 100 P\rs{ram}$ at the stream
border, where $P\rs{ram} = \rho u^2$ is the ram pressure. The 2D shock,
therefore, evolves similarly to the 1D overstable shock simulation,
alternating phases of expansion and collapse of the post-shock region
(see the on-line movie). The maximum thickness of the slab is $D\rs{slab}
\approx 1.4\times 10^9$ cm, i.e. less than in the 1D case by a
factor $\approx 1.8$. This result is analogous to that described by
\cite{2003ApJ...591..238S} in the case of 2D hydrodynamic radiative
shocks and is explained as due to the formation of denser knots of more
rapidly cooling gas. Note that, at variance with our MHD simulations,
the hydrodynamic model of \cite{2003ApJ...591..238S} does not include
the thermal conduction. In our simulations, the thermal conduction acts
as an additional cooling mechanism of the hot slab, draining energy from
the shock-heated plasma to the chromosphere, and partially contrasts the
radiative cooling. Depending on the temperature of the shock and on the
plasma $\beta$, therefore, the thermal conduction may lead to significant
differences between our results and those of \cite{2003ApJ...591..238S}.

\begin{figure}[!t]
  \centering
  \includegraphics[width=9.cm]{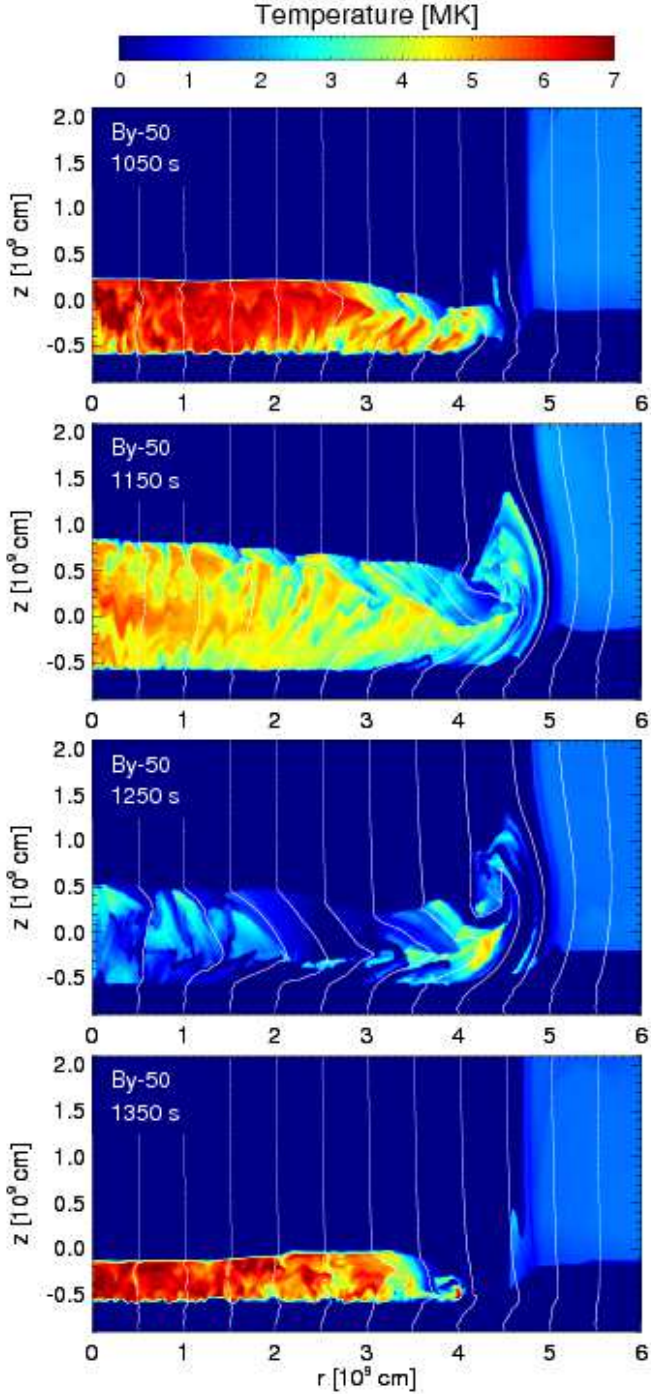}
  \caption{Temperature distribution in the $(r,z)$ plane, in linear
   scale, in the simulation By-50, at the labeled times. The white lines
   mark magnetic field lines.}
\label{fig6}
\end{figure}

Fig.~\ref{fig6} shows snapshots of the evolution of temperature
distribution (in linear scale) in run By-50. The post-shock region
gets larger ($t=1.05$~ks), reaches the maximum extension ($t=
1.15$~ks), and collapses ($t= 1.25$~ks); the cycle then repeats and
the slab begins to get larger again ($t= 1.35$~ks). Although the
magnetic field is strong enough to confine the post-shock plasma, the
plasma $\beta$ is slightly larger than 1 inside the slab and the 1D
approximation is not valid there. As in the 2D hydrodynamic simulations
(\citealt{2003ApJ...591..238S}), therefore, complex 2D cooling structures,
including knots and filaments of dense material, form there due to
the thermal instability of the post-shock plasma (see $t=1.15$~ks in
Fig.~\ref{fig6}). As discussed by \cite{2003ApJ...591..238S}, these 2D
complex structures lead to zones cooling more efficiently than those in 1D
models (by virtue of the increased cold-hot gas boundary) and, therefore,
the amplitude of oscillations is expected to be reduced. It is worth
noting that the similarity of our results with those of the hydrodynamic
models is due to the fact that, in run By-50, $\beta \gsim 1$ in the
slab and 2D cooling structures can form. In the case of $\beta \ll 1$
everywhere in the slab, we expect that the stream can be considered as
a bundle of independent fibrils (each of them describable in terms of
1D models), and the 2D MHD simulations would produce the same results
of 1D models.

We also study the global time evolution of 2D shocks by deriving
time-space plots of the temperature evolution analogous to that derived
for the 1D reference simulation (in order to compare directly our
1D and 2D results). From the 2D spatial distributions of temperature
and mass density, we first derive profiles of temperature along the
$z$-axis, by averaging the emission-measure-weighted temperature along
the $r$-axis for $r< 4\times 10^9$~cm (i.e. 80\% of the stream radius);
then, from these profiles, we derive the time-space plots of the average
temperature evolution.

\begin{figure}[!t]
  \centering
  \includegraphics[width=8.5cm]{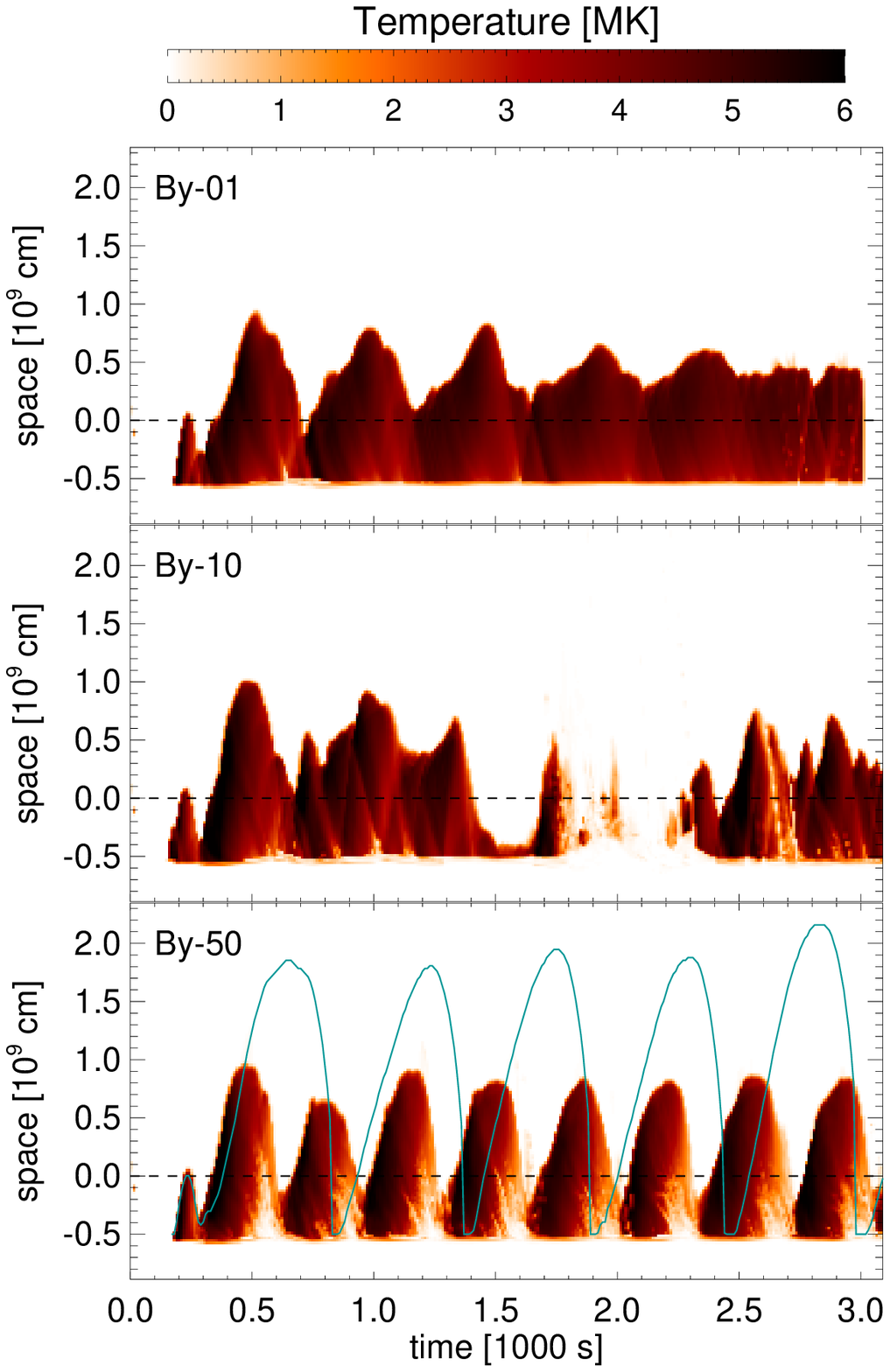}
  \caption{As in Fig.~\ref{fig3} for the 2D simulations By-01 (top),
   By-10 (middle) and By-50 (bottom). The blue curve in the bottom panel
   marks the shock position derived in the 1D reference model, HD-1D.}
  \label{fig7}
\end{figure}

Fig.~\ref{fig7} shows the results for runs By-01, By-10, and By-50, and
can be compared with Fig.~\ref{fig3}. In all the cases, the amplitude
of the shock oscillations is smaller than that observed in the 1D
reference model. In runs By-01 and By-10, the evolution of the 2D shock
is markedly different from that of the 1D shock (compare Fig.~\ref{fig3}
with top and middle panels in Fig.~\ref{fig7}): in run By-01, the
oscillations of the shock are stabilized after 2.5 ks and the solution
approaches a quasi-stationary state; in run By-10, the oscillations
appear chaotic without an evident periodicity (at least in the time
lapse explored here). As discussed before, the stabilization of the
shock oscillations in the former case is due to a small magnetic field
component perpendicular to the flow (\citealt{1993ApJ...413..176T}),
whereas the chaotic oscillations in the latter case are due to the
continuous stream perturbation by the material ejected sideways at the
stream base (see Fig~\ref{fig5}). In the run By-50, the shock evolves
somewhat similarly to the 1D overstable shock simulation, showing several
expansion and collapse complete cycles of the post-shock region (see
lower panel in Fig.~\ref{fig7}). However, as already discussed before,
some important differences arises: by comparing By-50 with HD-1D, the
oscillations are reduced in amplitude by a factor $\sim 1.8$ and occur
at higher frequency (period $P\rs{sh} \approx 300$~s).

\subsection{Effect of stream parameters}
\label{eff_param}

The details of the shock evolution described in this paper depend on
the model parameters adopted. In particular, the temperature of the
shock-heated plasma is determined by the free-fall velocity with which the
plasma impacts onto the star (Eq.\ref{temperat}); the stand-off height of
the hot slab generated by the impact depends on the velocity and density
of the stream (see Eq.~\ref{thickness}). Assuming a typical value for the
free-fall velocity of $400-500$ km s$^{-1}$ (leading to temperatures of
$\approx 3-5$ MK, as deduced from observations), the maximum thickness of
the slab is determined only by the density of the stream: the heavier the
stream, the thinner is expected the slab. Also the sinking of the stream
down in the chromosphere depend on the stream density and velocity at
impact. In fact, we find that the slab penetrates the chromosphere down
to the position at which the ram pressure, $\rho\rs{str0} u\rs{str0}^2$,
of the post-shock plasma equals the thermal pressure of the chromosphere
(see also \citealt{2008A&A...491L..17S}). As a result, heavier streams
are expected to sink more deeply into the chromosphere.

In the case of $\beta > 1$, the shock evolution is expected to be
influenced also by the stream radius, $r\rs{str}$. In fact, as shown
by our simulations, the complex plasma dynamics close to the border
of the stream (e.g. the generation of outflows of accreted plasma;
see runs By-01 and By-10) may affect the shock evolution in the inner
portion of the slab. The shock evolution is expected to be modified in
the whole slab if the oscillation period of the shock, $P\rs{sh}$, is
larger than the dynamical response time of the post-shock region, which
can be approximated as the sound crossing time of half slab: $\tau_{\rm
dyn} \sim r\rs{str}/c\rs{s}$, where $c\rs{s}$ is the isothermal sound
speed. In the cases discussed in this paper, the oscillation period
is $P\rs{sh}\approx 300$ s, and the sound crossing time of the slab is
$\tau_{\rm dyn}\approx 200$ s; as shown by our simulations, the shock
evolution is modified in the whole slab. Assuming an accretion stream
with twice the radius considered here, we find $\tau_{\rm dyn}\approx
400$ s, and the region at the center of the slab should not be affected
by the plasma dynamics at the stream border. In this case, we expect
quasi-periodic shock oscillations (analogous to those described by run
By-50) at the center of the stream and a strongly perturbed shock at
the stream border.

\subsection{Effect of spatial resolution}
\label{resol}

In problems involving radiative cooling, the spatial resolution and the
numerical diffusion play an important role in determining the accuracy
with which the dynamics of the system is described. In particular, in
the case of radiative shocks, we expect that the details of the plasma
radiative cooling depend on the numerical resolution: a higher resolution
may lead to different peak density and hence influence the cooling
efficiency of the gas, and therefore the amplitude and frequency of shock
oscillations.  In the simulations presented here, the thermal conduction
partially contrasts the radiative cooling alleviating the problem of
numerical resolution (see, for instance, \citealt{2008ApJ...678..274O}).

In order to check if our adopted resolution is sufficient to capture
the basic shock evolution over the time interval considered, we
repeated simulations By-01, By-10, and By-50 but with twice the spatial
resolution\footnote{Note that, in the case of simulations with higher
spatial resolution (very CPU time consuming), the time interval covered
was $\sim 1$ ks, instead of 3~ks, in order to reduce the computational
cost.} (runs By-01-HR, By-10-HR, and By-50-HR). The efficiency of
radiative cooling is expected to be the largest in the model with
$|\vec{B}\rs{0}|=50$~G, because there is no loss of material through
sideways outflows. In fact, quasi-periodic oscillations of the accretion
shock occur (see Fig.~\ref{fig7}) and make this case adequate for a
comparison of different spatial resolutions. Since this case is one of
the most demanding for resolution, it can be considered a worst case
comparison of convergence.

\begin{figure}[!t]
  \centering
  \includegraphics[width=9.cm]{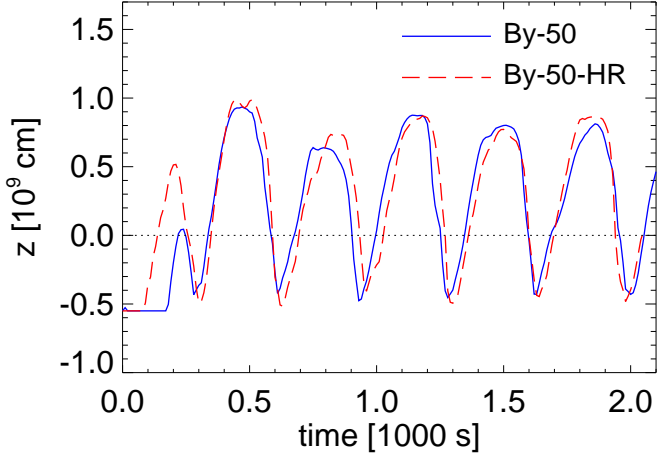}
  \caption{Evolution of the averaged shock-front location for runs By-50
   (solid blue line) and By-50-HR (dashed red line). The dotted line
   marks the initial position of the transition region between the
   chromosphere and the corona.}
\label{fig8}
\end{figure}

Fig.~\ref{fig8} compares the evolution of the averaged shock-front
location for runs By-50 and By-50-HR. The main differences between the
simulations appear in the first bounce which starts earlier and is larger
by a factor $\approx 2$ in By-50-HR than in By-50. The first expansion of
the hot slab occurs after that the stream has penetrated the chromosphere
down to the position where the ram pressure of the shock-heated material
equals the thermal pressure of the chromosphere. The first bounce is a
transient, therefore, and in fact its amplitude and width are different
from those of subsequent bounces in all the simulations examined. On
the other hand, Fig.~\ref{fig8} clearly shows that, apart from the first
bounce, in general, the results of the two simulations agree quite well,
showing differences $< 10\%$. We expect, therefore, that increasing the
spatial resolution, the results of our simulations may slightly change
quantitatively (e.g. the amplitude of oscillations) but not qualitatively.

\section{Discussion}
\label{sec4}

\subsection{Distribution of emission measure vs. temperature of
the shock-heated plasma}
\label{emt}

As discussed in the introduction, there is a growing consensus that
two distinct plasma components contribute to the X-ray emission of
CTTSs: the stellar corona and the accretion shocks. This idea has been
challenged recently by \cite{2009A&A...507..939A}, who compared the
distributions of emission measure, EM$(T)$, of two CTTSs with evidence
of X-ray emitting dense plasma (MP~Mus and TW~Hya) with that of a star
(TWA 5) with no evidence of accretion (i.e. only the coronal component
is present) and with the EM($T$) derived from a 1D hydrodynamic model
of accretion shocks (i.e. only the shock-heated plasma component is
present; \citealt{2008A&A...491L..17S}). They proved that the EM($T$)
of MP~Mus and TW~Hya can be naturally interpreted as due to a coronal
component, dominating at temperatures $T> 5$ MK, plus a shock-heated
plasma component, dominating at $T < 5$ MK.

In case of shocks with $\beta > 1$, our simulations show that the
distributions of temperature and density at the base of the accretion
column can be rather complex (see, for instance, Fig.~\ref{fig6}). It is
interesting, therefore, to investigate how the EM$(T)$ of the shock-heated
plasma changes with $\beta$ and if it is possible to derive a diagnostics
of the plasma-$\beta$ in the post-shock region.

In order to derive the EM($T$) distribution of the accretion region from
the models, we first recover the 3D spatial distributions of density
and temperature by rotating the corresponding 2D distributions around
the symmetry $z$ axis ($r=0$). The emission measure in the $j$th
domain cell is calculated as $\em_{\rm j} = n_{\rm Hj}^2 V_{\rm j}$,
where $n_{\rm Hj}^2$ is the particle number density in the cell, and
$V_{\rm j}$ is the cell volume. The EM($T$) distribution is then derived
by binning the emission measure values into slots of temperature; the
range of temperature [$5 < \log T (\mbox{K}) < 8$] is divided into 15
bins, all equal on a logarithmic scale ($\Delta \log T = 0.2$).

\begin{figure}[!t]
  \centering
  \includegraphics[width=8.5cm]{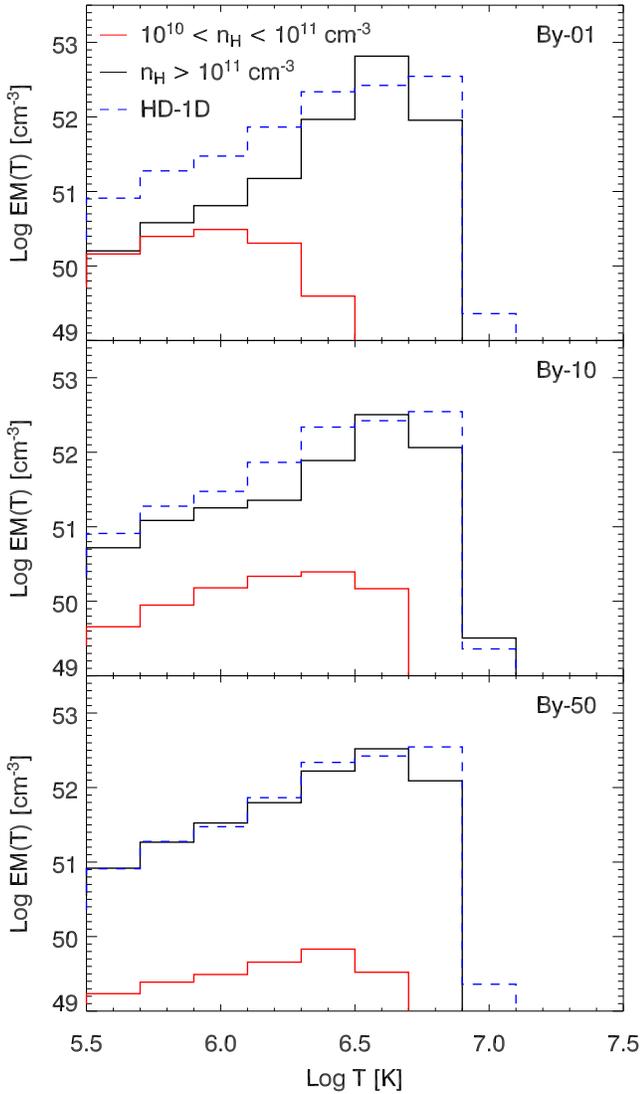}
  \caption{Distributions of emission measure vs. temperature, averaged
           over 3 ks, for runs By-01, By-10, and By-50. Black (red)
           lines mark the average EM($T$) distributions of plasma with
           density $n\rs{H} > 10^{11}$ cm$^{-3}$ ($10^{10} < n\rs{H}
           < 10^{11}$ cm$^{-3}$); dashed blue lines mark the average
           EM($T$) distribution for the 1D reference model HD-1D.}
  \label{fig9}
\end{figure}

Fig.~\ref{fig9} shows the EM($T$) distributions averaged over 3 ks
for runs By-01, By-10, and By-50, together with the average EM$(T)$
derived from our 1D reference model HD-1D (blue dashed lines). The
figure shows the EM($T$) distributions of plasma with density $n\rs{H} >
10^{11}$ cm$^{-3}$ (black) and $10^{10} < n\rs{H} < 10^{11}$ cm$^{-3}$
(red). The 2D MHD models have been normalized in order to have an
X-ray luminosity $L\rs{X}\approx 4\times 10^{29}$ erg in the band
$[0.5-8.0]$ keV, in agreement with the luminosity derived from the
low-temperature ($\log T < 6.7$) portion of the EM($T$) distribution
of MP~Mus (\citealt{2009A&A...507..939A}). This is obtained assuming
that $\approx 10$ accretion streams similar to that modeled here are
present simultaneously. The 1D model has been normalized in order to
match the EM peak in By-50. Inspecting Fig.~\ref{fig9} we note that:
i) in all the cases, the EM($T$) has a peak at $\sim 5$ MK and a shape
compatible with those observed in MP~Mus and TW~Hya and attributed to
shock-heated material (e.g. \citealt{2009A&A...507..939A}); ii) most
of the X-ray emission arises from shock-heated plasma with density
$> 10^{11}$ cm$^{-3}$ regardless of the $\beta$ (i.e. the most dense
component of the post-shock region dominates); iii) the slope of the
ascending branch of the EM($T$) distribution is comparable in runs
By-50 and HD-1D, and gets steeper for decreasing values of $\beta$. The
time-averaged X-ray luminosity\footnote{The synthetic X-ray spectra in
the band $[0.5-8.0]$ keV have been derived with the PINTofALE spectral
code \citep{Kashyap2000BASI} with the APED V1.3 atomic line database
\citep{Smith2001ApJ}, assuming the metal abundances adopted in
the whole paper, namely 0.5.} derived from these EM($T$) distributions
ranges between $5\times 10^{29}$ erg (run By-10) and $8\times 10^{29}$
erg (run By-01), having the X-ray luminosity only a weak dependence on
the plasma $\beta$.

Our model supports, therefore, the idea that dense shock-heated plasma may
contribute significantly to the low temperature portion of the EM($T$)
distributions of CTTSs regardless of the $\beta$ value. We also suggest
that the shape of the EM($T$) might be used as a diagnostics of $\beta$ in
the post-shock region, if the coronal contribution to the low temperature
tail can be neglected.

\subsection{Mass accretion rates}
\label{accr_discussion}

Time-dependent models of radiative accretion shocks provide a strong
theoretical support to the hypothesis that soft X-ray emission from CTTSs
arises from shocks due to the impact of the accretion columns onto the
stellar surface. However, several points remain still unclear. Among
these, the most puzzling is probably the fact that the mass accretion
rates derived from X-rays, \rate$\rs{X}$, are consistently lower by one
or more orders of magnitude than the corresponding {\rate} values derived
from UV/optical/NIR observations (e.g. \citealt{2005ESASP.560..519D,
Schmitt2005A&A, Gunther2007A&A, 2009A&A...507..939A}, Curran et al.  2009,
in preparation). We have here the opportunity to discuss the problem
of the accretion rate, in particular focusing on the cases with $\beta
\gsim 1$ in the post-shock region.

The model parameters adopted in this paper describe a stream with an
accretion rate of $\approx 9\times 10^{-12} M\rs{\odot}$~yr$^{-1}$.
According with the discussion in Sect.~\ref{emt}, $\approx 10$ streams
are needed to match the soft X-ray luminosity of MP~Mus. In this case
the accretion rate is $\approx 9\times 10^{-11} M\rs{\odot}$~yr$^{-1}$,
which is in nice agreement with that deduced from observations,
namely \rate$\rs{X} = 5-8\times 10^{-11} M\rs{\odot}$~yr$^{-1}$
(\citealt{Argiroffi2007A&A}). Alternatively, the observed \rate$\rs{X}$
may be reproduced by our model if the accretion stream has a larger cross
section (with radius $r\rs{str} \approx 10^{10}$ cm). In this case,
we expect some changes to the dynamics of the shock-heated plasma as
described in Sect.~\ref{eff_param}.

On the other hand, the mass accretion rate of MP~Mus, as deduced
from optical observations, is \rate$\rs{opt} \approx 3\times
10^{-9}M\rs{\odot}$~yr$^{-1}$ (\citealt{2009A&A...507..939A}),
exceeding by more than one order of magnitude the value obtained from
X-rays. Analogous discrepancies are found in all CTTSs for which it
is possible to derive \rate$\rs{X}$ (see also Curran et al. 2009, in
preparation). The discrepancy might be reconciled if \rate$\rs{X}$ values
are underestimated due, for instance, to absorption from optically thick
plasma. However, as we explain below, even assuming that the absorption
can account for the observed {\rate} discrepancy, the idea that the
same streams determine both \rate$\rs{opt}$ and \rate$\rs{X}$ has to be
discarded. In fact, assuming the accretion parameters adopted here, we
derive that $\approx 300$ streams must be present to match \rate$\rs{opt}$
or the stream should have a cross section $\approx 2\times 10^{22}$
cm$^2$, implying that, in both cases, $\approx 20$\,\% of the stellar
surface should be involved in accretion. None of these hypothesis is
realistic, being the surface filling factor of hot spots resulting from
accretion up to a few percent (e.g.  \citealt{1989A&A...211...99B,
1990ApJ...349..190H, 1993A&A...272..176B, 1994A&A...287..131G,
1994AJ....107.2153K, 2009ApJ...703.1224D} and references therein).

The above inconsistency can be removed if the mass density of accretion
streams is larger than assumed here; for instance, for a density of
the stream $n_{\rm str0} = 5\times 10^{12}$ cm$^{-3}$ (a factor 50
larger than modeled), the mass accretion rate is $\approx 5\times
10^{-10}M\rs{\odot}$~yr$^{-1}$ and few streams are needed to match
\rate$\rs{opt}$. In this case, however, the model cannot explain the
lower values of density derived from X-ray observations ($n_{\rm str0}
\approx 10^{11}$ cm$^{-3}$; \citealt{Argiroffi2007A&A}).

A possible solution to remove the discrepancy in the {\rate} values
(and in the $n_{\rm str0}$ values) is that few accretion streams
characterized by different mass density values are present at the same
time: those with density $\approx 10^{11}$ cm$^{-3}$ would produce
shocks that are visible in the X-ray band, leading to the observed
\rate$\rs{X}$; those with higher densities would produce shocks not
visible in the X-ray band and leading to the observed \rate$\rs{opt}$
(being the dominant component in the UV/optical/NIR bands). The reason
of the different visibility could be due to local absorption of the
X-ray emission, as explained in the following.

Our simulations show that the accretion stream penetrates the chromosphere
down to the position at which the ram pressure of the post-shock
plasma equals the thermal pressure of the chromosphere (see also
\citealt{2008A&A...491L..17S}). Part of the shock-heated plasma is buried
down in the chromosphere and is expected to be obscured by significant
absorption from optically thick plasma (see also Sacco et al. 2009, in
preparation). In the simulations presented here, this portion is 1/3 of
the hot slab ($h\rs{sink} \approx 5\times 10^8$ cm; see Fig.~\ref{fig7}),
and most of the post-shock plasma (above the chromosphere) is expected
to be visible with minimum absorption. Instead, in denser streams,
the shock column is buried more deeply in the chromosphere, due
to the larger ram pressure, and its maximum thickness is smaller,
according to Eq.~\ref{thickness} (see Sect. \ref{eff_param}). Assuming
the accretion parameters adopted in this paper but with $n_{\rm str0}
= 5\times 10^{12}$ cm$^{-3}$, Eq.~\ref{thickness} gives $D\rs{slab}
\approx 5\times 10^7$ cm, that is much smaller than the expected sinking
of the stream in the chromosphere ($h\rs{sink} \approx 7\times 10^8$
cm). In this case, the post-shock column is buried under a hydrogen
column density $N\rs{H} = n\rs{H} h\rs{sink} \approx 2\times 10^{22}$
cm$^{-2}$ and the photoelectric absorption of the O{\sc vii} triplet
(i.e. the lines commonly used to trace the accretion in the X-ray band)
is given by $\exp[-\sigma\rs{OVII}N\rs{H}]\approx 10^{-6}$. We conclude,
therefore, that heavy streams may produce X-ray emitting shocks that,
however, are hardly visible in X-rays, being buried too deeply in the
chromosphere.

\subsection{Variability of X-ray emission from shock-heated plasma}

Another important point in the study of accretion shocks in CTTSs is
the periodic variability of X-ray emission, due to the quasi-periodic
oscillations of the shock position induced by cooling, predicted by
time-dependent 1D models. However, in the only case analyzed up to date,
namely TW~Hya, no evidence of periodic variations of soft X-ray emission
(thought to arise predominantly in an accretion shock) has been found
(\citealt{2009ApJ...703.1224D}). This result apparently contradicts the
prediction of current 1D models and Drake et al. suggested that these
models might be too simple to explain the 3D shock structure.

On the other hand, quasi-periodic shock oscillations are
expected if the accretion stream is homogeneous and constant
(no variations of mass density and velocity). However, there is
substantial observational evidence that the streams are clumped and
inhomogeneous (e.g. \citealt{1996A&A...307..791G, 1998ApJ...494..336S,
2003A&A...409..169B, 2007A&A...463.1017B}). In these conditions, periodic
shock oscillations are expected to be hardly observable.

The 2D MHD simulations presented here show that, even assuming constant
stream parameters, periodic oscillations are not expected if $\beta \gg 1$
in the post-shock region (runs By-01 and By-10). In these cases, in fact,
the time-space plots of temperature evolution (top and middle panels in
Fig.~\ref{fig7}) predict that the oscillations may be rapidly dumped,
approaching a quasi-stationary state with no significant variations of
the shock position, or the variability may be chaotic (with no obvious
periodicity) due to strong perturbation of the stream by the accreted
material ejected sideways. In none of these cases, therefore, we expect
to observe periodic modulation in the X-ray emission.

On the other extreme, for shocks with $\beta \ll 1$, we expect that the
single accretion stream is structured in several fibrils, each independent
from the others due to the strong magnetic field which prevents mass and
energy exchange across magnetic field lines. Time-dependent 1D models
describe one of these fibrils. Being independent from each other, the
fibrils can be characterized by slightly different mass density and
velocity, which would result in different instability periods, as well
as by random phases of the oscillations. \cite{2009ApJ...703.1224D},
assuming that 1D models describe a single stream, derived the number
of streams needed to account for the absence of periodic variability
in TW~Hya and concluded that this number contrasts with the presence
of conspicuous rotationally modulated surface flux with small filling
factor. Following \cite{2009ApJ...703.1224D} but considering the fibrils
instead of the streams, we may argue that an accretion stream consisting
of $200-300$ different fibrils with radius $r\rs{fibr}\approx 10^8$ cm
and with different instability periods and random phases would produce
a signal pulsed at a level of less than 5\% as measured in TW~Hya.

The intermediate situation is for shocks with $\beta$ around 1. In this
case, our run By-50 shows quasi-periodic oscillations of the shock
position (see bottom panel in Fig.~\ref{fig7}) and predicts periodic
variations of the X-ray emission arising from a single stream. In
fact, in this case, the magnetic field is strong enough to confine
the shock-heated plasma, but it is too weak to consider valid the 1D
approximation inside the slab that cannot be described as a bundle of
fibrils. As a result, the shock oscillates coherently in the slab. As
noted by \cite{2009ApJ...703.1224D}, in this case it is not possible to
reproduce the absence of periodic modulation of X-ray emission observed
in TW~Hya, and we conclude that shocks with $\beta \approx 1$ do not
occur in this star.

It is worth emphasizing that further investigation is needed to
understand how much the intermediate case described by run By-50 is
frequent and observable: an intensive simulation campaign is needed to
assess the range of $\beta$ values leading to streams with quasi-periodic
oscillations; also a systematic analysis of the variability of soft
X-ray emission should be performed, considering a complete sample of
CTTSs with evidence of X-ray emitting accretion shocks, to assess if
and when periodic variability is observed.

\subsection{Effects of accretion shocks on the surrounding stellar
atmosphere}

In the case of shocks with $\beta \gg 1$, our simulations predict that
the stellar atmosphere around the region of impact of the stream can be
strongly perturbed by the impact, leading to the generation of MHD waves
and plasma motion parallel to the stellar surface: the larger is the
plasma $\beta$ in the post-shock region, the larger is the perturbation
of the atmosphere around the shocked slab. The resulting ejected flow
also advects the weak magnetic field such that \referee{the conditions
for ideal MHD may break down, magnetic reconnection may be possible
and eventually releases the stored energy from the magnetic field} (not
described however by our model that does not include resistivity effects).

A possible effect of the perturbation of the stellar atmosphere is that
shocks with $\beta \gg 1$ may contribute to the stellar outflow. In fact,
\cite{2008ApJ...689..316C} suggested that the MHD waves and the material
ejected from the stream (as in our runs By-01 and By-10) may trigger
stellar outflow and proposed a theoretical model of accretion-driven winds
in CTTSs (see, also, \citealt{2009ApJ...706..824C}). His model originates
from a description of the coronal heating and wind acceleration in the
Sun and includes a source of wave energy driven by the impact of accretion
streams onto the stellar surface (in addition to the convection-driven
MHD turbulence which dominates in the solar case). The author found that
this added energy seems to be enough to produce T Tauri-like mass loss
rates. It would be interesting to assess how the different plasma-$\beta$
cases discussed in this paper contribute to the added energy.

\subsection{Limits of the model}

Our simulations were carried out in 2D cylindrical geometry, implying
that all quantities are cyclic on the coordinate $\phi$. This choice
is expected to affect some details of the simulations but not our
main conclusions. In particular, adopting a 3D Cartesian geometry,
the simulations would provide an additional degree of freedom for
hydrodynamic and thermal instabilities, increasing the complexity of
cooling zones and, therefore, the cooling efficiency of the plasma in
the post-shock region. As a result, in 3D simulations, the amplitude
of the shock oscillations might be slightly smaller and the frequency
slightly higher than that observed in 2D simulations.

As discussed in Sect. \ref{eff_param}, some details of our simulations
depend on the choice of the model parameters. For instance, the
temperature of shock-heated plasma, the stand-off height of the hot slab
as well as the sinking of the stream down in the chromosphere depend
on the stream density and velocity at impact. The accretion parameters
adopted here originate from the values derived by \cite{Argiroffi2007A&A}
for MP~Mus; the cases presented here are representative of a regime in
which the shock-heated plasma has a temperature $\approx 5$ MK and part
of the slab is above the chromosphere, making it observable with minimum
absorption. The range of magnetic field strength considered has been
chosen in order to explore shocks with $\beta \gsim 1$. Nevertheless,
our results undoubtedly show that the stability and dynamics of accretion
shocks strongly depends on $\beta$ and that a variety of phenomena,
not described by 1D models, arises, including the generation of plasma
motion parallel to the stellar surface and MHD waves.

It is worth noting that our model does not include magnetic resistivity
effects. The complex shock evolution described by our simulations
show that violent plasma flows may advect the magnetic field such
that \referee{the conditions for ideal MHD may break down and magnetic
reconnection may possibly} occur (not described by our simulations). In
particular, field lines at the stream border are squeezed close together
and may reconnect, resulting in a change of the magnetic topology as
well as in a release of magnetic energy. Since most of these events are
expected to occur at the stream border, magnetic reconnection may play
an important role in the dynamics and energetic of the ejected accreted
plasma. For instance, the magnetic reconnection may trigger flares at
the stream border outside the accretion column. This scenario is not in
contradiction with \cite{2009pjc..book..179R}, who showed that it is
unlikely that flares can be triggered in an accreting flux tube. This
issue deserves further investigation in future studies.

\section{Conclusion}
\label{sec5}

We investigated the stability and dynamics of accretion shocks in CTTSs,
considering the case of $\beta \gsim 1$ in the post-shock region,
through numerical MHD simulations. To our knowledge, the simulations
presented here represent the first attempt to model 2D accretion
shocks that simultaneously includes magnetic fields, radiative cooling,
and magnetic-field-oriented thermal conduction. Our findings lead to
several conclusions:

\begin{enumerate}

\item In all the cases, a hot slab of shock-heated material is
generated at the base of the accretion column due to the impact of the
stream with the chromosphere. In the case of shocks with $\beta > 10$,
violent outflows of shock-heated material and, possibly of MHD waves,
are generated at the border of the hot slab and they may perturb the
surrounding stellar atmosphere. For shocks with $\beta\approx 1$, the
shock-heated plasma is efficiently confined by the magnetic field and
no outflow forms.

\item If the magnetic field is too weak to confine the shock-heated
plasma but is strong enough to keep it close to the stream, the escaped
accreted material may strongly perturb the accretion column, modifying
the dynamics and stability of the shock itself.

\item The accretion shocks are susceptible to radiative shock
instability. The resulting shock oscillations strongly depend on the
plasma $\beta$: for $\beta > 10$, the oscillations may be rapidly
dumped by the magnetic field, approaching a quasi-stationary state,
or may be chaotic with no obvious periodicity due to perturbation of
the stream induced by the post-shock plasma itself; for $\beta$ around 1
the oscillations are quasi-periodic with amplitude smaller and frequency
higher than those predicted by 1D models.

\end{enumerate}

We also suggest that the {\rate} discrepancy observed in CTTSs may
be solved if few streams with significantly different mass density
are present: heavy streams are not visible in X-rays due to strong
absorption and determine the {\rate} value deduced from observations
in UV/optical/NIR bands; light streams can be observed in X-rays and
determine the {\rate} value deduced from X-ray observations. As for the
absence of periodic X-ray modulation due to shock oscillations found in
TW~Hya by \cite{2009ApJ...703.1224D}, we find that no periodic modulation
of X-rays is expected in cases of shocks with either $\beta \ll 1$ or
$\beta \gg 1$, whereas periodic modulation is found for $\beta$ around 1,
indeed a rather limited set of possible cases. We interpret the absence
of periodic X-ray modulation in TW~Hya as evidence of the fact that, in
this star, $\beta$ differs significantly from 1 in the accretion shocks
or the streams are inhomogeneous.

\begin{acknowledgements}
We thank the referee for constructive and helpful criticism. We are
grateful to Andrea Mignone and Titos Matsakos for their support in
using the \PLUTO\ code. \PLUTO\ is developed at the Turin Astronomical
Observatory in collaboration with the Department of General Physics of the
Turin University. The simulations have been executed at CINECA (Bologna,
Italy), and at the HPC facility (SCAN) of the Osservatorio Astronomico di
Palermo. This work was supported in part by the EU Marie Curie Transfer
of Knowledge program PHOENIX under contract No. MTKD-CT-2005-029768 and
by Agenzia Spaziale Italiana under contract No. ASI-INAF I/088/06/0.
\end{acknowledgements}

\bibliographystyle{aa}
\bibliography{biblio}

\end{document}